# APPROACHES TO CURRICULUM AND TEACHING MATERIALS TO BRING OUT BETTER SKILLED SOFTWARE ENGINEERS – AN INDIAN PERSPECTIVE

**H. A. Padmini**

Infosys Technologies Ltd Bangalore,
Karnataka, India -560100
padmini_ha@infosys.com

**A. Keshav Bharadwaj**

**T. R. Gopalakrishnan Nair**

RIIC, Dayananda Sagar Institutions, DSCE
Bangalore , Karnataka, India - 560078
akeshavb@hotmail.com

trgnair@yahoo.com

## Abstract

Development of Curriculum and delivery materials has undergone changes over a period of time, in undergraduate engineering degree system in Indian universities. However, there exists a gap between industry expectations in IT field and skills and knowledge that the graduating engineers possess and this continues to grow. A similar situation has been seen in the developed countries like USA, UK and Australia. Several researchers and practitioners have discussed and tried to come up with innovative approaches to teaching software engineering and IT as a whole. In India, it is of vital importance that steps be taken to address this issue seriously. This paper discusses some of the measures that have been implemented so that this gap is reduced and software engineers with better skills are produced. Changes to curriculum, industry-academia collaboration through conferences, sabbaticals etc., industry internships and live projects for final year students are some of the measures that have been discussed in this paper. The implementation of these measures may lead to fulfilling the growing requirement for skilled software engineers who can handle the industry challenges.

## 1 INTRODUCTION

Rapid industrialization and economic growth has made India a force to reckon with in the world of technology. India has also made remarkable progress in engineering and technical education since independence. Undergraduate engineering education is imparted in the numerous institutions across the country and numerous graduates are produced every year. Software engineering as a field of practice emerged about three decades ago and relevant graduate and post graduate engineering courses were introduced in the country's institutions. Two main streams with different organizational setup viz., Computer Science and Information Science now produce software engineering graduates. These in addition to other engineering graduates cater to the requirements of the software industry.

This paper analyzes the existing curriculum and discusses the teaching content and practices across universities and industry (for the Computer science and Information Science courses). Several changes have already been introduced and many more are needed. This would ensure that the graduates who pass out with degrees in the above mentioned two engineering courses are better equipped to handle the challenges of the software industry. Academia-Industry relationships, existing barriers, modes of overcoming the barriers and other contentious issues have been detailed in [7]

This paper is organized as follows: Section 2 discusses the current curriculum in a sample of institutions while Section 3 discusses about the courses that are covered in initial training given by software companies to their newly recruited employees. This is followed by Section 4 which discusses remedial measures that are needed/have been taken at universities and institutions. Section 5 details some remedial measures taken/that are needed by corporates to help universities/ institutions prepare better qualified software engineers and is followed by conclusions in Section 6.

## 2 CURRENT CURRICULUM OF COMPUTER SCIENCE AND INFORMATION SCIENCE COURSES

When graduate courses in Computer Science was introduced about two decades ago (and more recently when Information Science degree was introduced) the curriculum was discussed at length and was developed by distinguished persons working in this field based on similar courses being conducted in countries like USA, UK, etc. Over the years some changes have been introduced in the syllabus and the subjects covered. The subjects which were earlier part of the curriculum and those that have now been introduced are shown in the table below.

**Table I – Snapshot of earlier and present subjects in Curriculum for various Semesters of the 4- yr Engineering Degree [I]**

| | |
|---|---|
| **Subjects which were part of curriculum earlier** | Network Analysis, Computer Programming in Basic and Fortran, Analog Electronics and Analog Computer, Computational Methodology, Structured Programming and Pascal, Cobol and Business Application, Control Systems, Probability, Statistics and Queuing, Digital Electronics, Professional Communication and Report Writing, Principles of Programming languages, and Computer Architecture. |
| **Subjects which have been introduced later** | Object Oriented Programming with C++, Software Engineering, Systems Software, Object Oriented Modeling and Design, Programming the Web, Data Mining, Software Testing etc. |

Some subjects like Basics of Computers, Computer programming in Visual Basic, Structured programming and C/C++, Probability and Statistics are now covered by students in their pre-university degree or school curriculum itself.

The present day syllabus for the various semesters is shown in the table given below. This is the curriculum used in some institutions in India for Information Science and Engineering. The subjects covered in the 1st and 2nd semesters are common across disciplines and deal with subjects like mathematics, physics, chemistry, engineering graphics and drawing, etc.

**Table 2 – Curriculum for III-VIII Semesters of the 4- yr Engineering Degree [I]**

| **Semester** | **Subjects** |
|---|---|
| III Semester | Mathematics , Electronic Circuits, Logic Design, Discrete Mathematical Structures, Data Structures with C, Unix and Shell programming, Data Structures Laboratory, Electronic Circuits and Logic Design Laboratory. |
| IV Semester | Mathematics, Graph Theory and Combinatorics, Analysis and Design of Algorithms, Object Oriented Programming with C++, Microprocessors, Computer Organization, Object Oriented Programming Laboratory, Microprocessors Laboratory. |
| V Semester | Software Engineering, Systems Software, Operating Systems, Database Management Systems, Computer Networks –I, Formal Languages and Automata Theory, Database Applications Laboratory, Algorithms Laboratory. |
| VI Semester | Management and Entrepreneurship, Unix System Programming, File Structures, Computer Networks – II, Information Systems, File Structures Laboratory, Systems Software Laboratory. |
| VII Semester | Object Oriented Modeling and Design, Software Architectures, Programming the Web, Data Mining, Networks Laboratory, Web Programming Laboratory. |
| VIII Semester | Software Testing, System Modeling and Simulation, Project Work, Seminar. |

## 3 SUBJECTS/TOPICS WHICH SOFTWARE COMPANIES INCLUDE AS PART OF THEIR INDUCTION TRAINING FOR SOFTWARE ENGINEERS

Most software corporates have their newly recruited engineers (generally from Computer Science and Information Science streams) undergo three to six months training during which some new subjects are covered and other previously learnt concepts are reinforced. Some of the subjects covered in the training include the following:[2]

**Table 3 – Subjects covered during the training given to new employees [2]**

| Technical training | Soft skills training |
|---|---|
| Analysis of Algorithms, C and/or C++, Computer hardware and System Software Concepts, Object oriented concepts, Programming Fundamentals, Systems Development Methodology, Relational DBMS, Java, Client Server Concepts, Orientation and Problem Solving, System Development Methodologies, User Interface Design. | Verbal Assessment, Interpersonal communication, Working in teams, Customer Orientation, English, Communication using e-mail etc. |

This clearly shows that there is some repetition of technical subjects which though helpful in reinforcement of concepts may sometimes bring a negative reaction from the new employees by making them bored and uninterested. It is also observed that despite such induction training given to new employees they are unable to manage the challenges of projects. Training should be on-going throughout the lifetime of a person. This arrangement has time and cost implications for the organization. Hence it is essential to determine the changes needed to produce effective software engineers.

## 4 REMEDIAL AND INNOVATIVE MEASURES TO BE INTRODUCED AT THE UNIVERSITY LEVEL:

### 4.1 Changes to Curriculum and teaching materials to keep them abreast with technology changes

It is of vital importance to "introduce software engineering research and industry best practices into the curriculum" [8]. Rapid advances in technology require similar changes to be mirrored in the curriculum and teaching materials of universities and institutions. It is necessary to not only follow but also sometimes to take the lead by including seminal research areas as subjects in Engineering. For example Object Oriented concepts were introduced in the curriculum about 10-15 years after it had been established in the industry. Now Agile methodology is yet to be added with the Waterfall methodology which exists in the curriculum. Similarly recent researches in Web services, Semantic Web etc need to be introduced. Leading and big software companies must help align the college curriculum with the industry's requirements and work with educational bodies towards implementing it [2].Presently technologies which are new are rarely introduced in the curriculum. Only after the technology has been in use for 10-15 years is it considered for inclusion. It could be worth the while to analyze upcoming technologies and consider them positively when curriculum changes are being discussed.

In many autonomous universities curriculum has been constantly changing to reflect technology. The non-autonomous institutes must cut out the red-tapism and bureaucracy and introduce necessary changes to curriculum and teaching materials. A guideline such that changes in syllabus are made compulsorily every four years should be introduced. Also experts from the industry should be on the board to oversee the syllabus changes.

### 4.2 Industry–academia collaboration through seminars, conferences, lectures and sabbaticals

Software Engineering Practitioners must be responsive to the societal needs and be willing to share their experiences, knowledge, and expertise with students by giving lectures, holding workshops and interacting with them on a one-to-one basis wherever possible. Companies must help organize seminars and training sessions particularly for the faculty to give them an industry perspective, enabling them to train the students accordingly. Industry-oriented topics must be determined, courseware provided to students; and projects given to them. Institutions and Universities must also provide sabbaticals for the faculty wherein the faculty is involved in projects in different companies for 3-6 months and are able to obtain hands on experience. This is very important as the experience that the faculty gains here is further transmitted to the students [2]. Several companies like Infosys Technologies Ltd, Wipro Ltd, IBM sponsor conferences and seminars and also collaborate with the academia by offering 3-6 month project work to faculty. An obstacle to implementing these can be that the cost and time spent by the company might be considered wastage by them. Actually companies can use this opportunity for brand building and attracting the cream of students into their companies. Companies that have participated in such academic collaborations have gained considerably from these practices.

### 4.3 Training in Soft Skills to both faculty and students

Programmes aiming at grooming individuals into excellent team players, with strong analytical, interpersonal and communication skills, will help students to adapt to the corporate work culture easily. For e.g. good communication skills are necessary in addition to strong technical skills to work comfortably with peers. Hence training in soft skills must be provided to faculty and students preferably with or without the co-operation of software companies. Corporate training institutes can fill the gap and provide soft skill courses. Universities can provide grants to individual faculty members to avail training in soft skills.

### 4.4 Professional and Social Responsibility

Students graduating from colleges always face the dilemma – what is more important – ethics or success? A course on Business and Professional Ethics in the final semester would groom the fresh engineers and set their expectation as to what behaviour is acceptable and what is not. An employee is more effective only when he/she follows the values set by the organization.

### 4.5 Cultural Orientation

Going forward, a lot of people from Tier II and Tier III cities would be migrating to the metros. Courses on diversity, cultural sensitivity, women at work, challenges in a big city, work life balance would equip the students to face the real life challenges. With more and more multinationals setting up shop in India, they are attracting the best students due to the fat salaries they are offering. In these companies, the employees may be posted out of India and hence have to interact with foreigners. Hence a grooming course as to how they should behave should be provided to them.

### 4.6 Faculty Empowerment

The college administration should empower the faculty members and involve them in decision making. They should be encouraged to tailor new programs that are relevant and this practice would give them a sense of ownership.

## 5 REMEDIAL AND INNOVATIVE MEASURES TO BE INTRODUCED BY INDUSTRY:

### 5.1 Industry contribution

IBM has an initiative “Drona”, wherein the employees advise the lecturers of colleges what courses are relevant to graduating engineers and train them in those courses. The objective of this initiative is to make the graduating engineers “employable” right from day one. IBM also has a programme called “Career Education in IBM Software” (CEIS) which brings together existing IBM Software Education courses into a simple and coherent program targeted for specific career paths. The CEIS program prepares employees to face the demands of the present day software career. The training incorporates a “blended learning” approach. It integrates classroom training, lab sessions and live projects thus providing students theoretical and practical training [3].

Infosys Technologies Limited has an initiative “Campus Connect” to enhance the quality and quantity of the IT talent-pool and sustain the growth of the IT industry itself. Some key programmes under this initiative are the “Faculty Enablement Programme” , “Soft Skills Programme”, “Sabbaticals for Faculty” etc which introduce to the engineering college faculty generic and soft skill courses aiming at a paradigm shift in the learning and teaching methodology at an accelerated pace, to increase effectiveness.

The way the faculty train the students, assess them and put technology to use undergoes a major change. It exposes faculty to the Infosys way of professional training and sharing industry best practices [2]. The figures below show the model and content of the “Campus Connect” program.

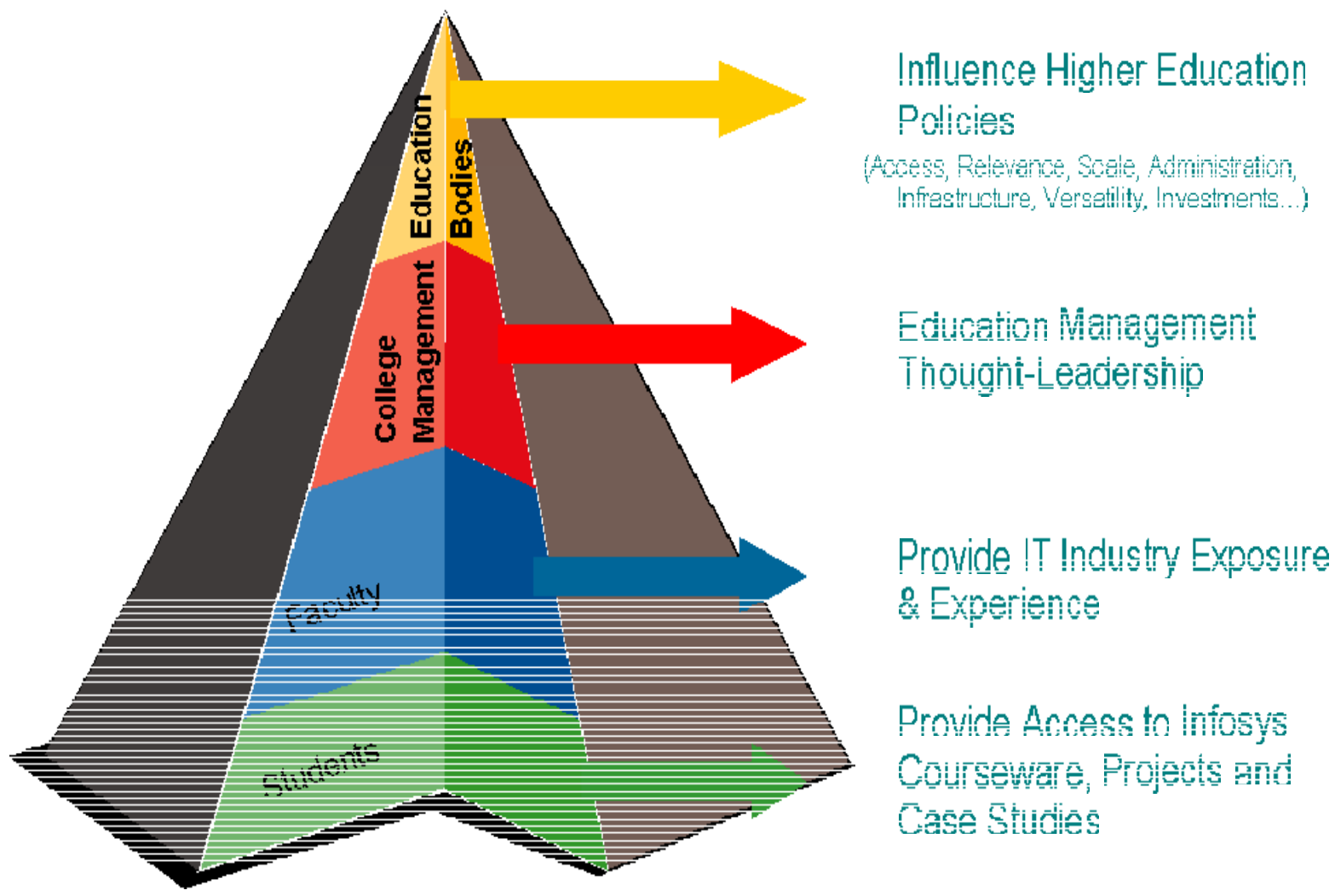

**Fig 1 – Programme Approach [2]**

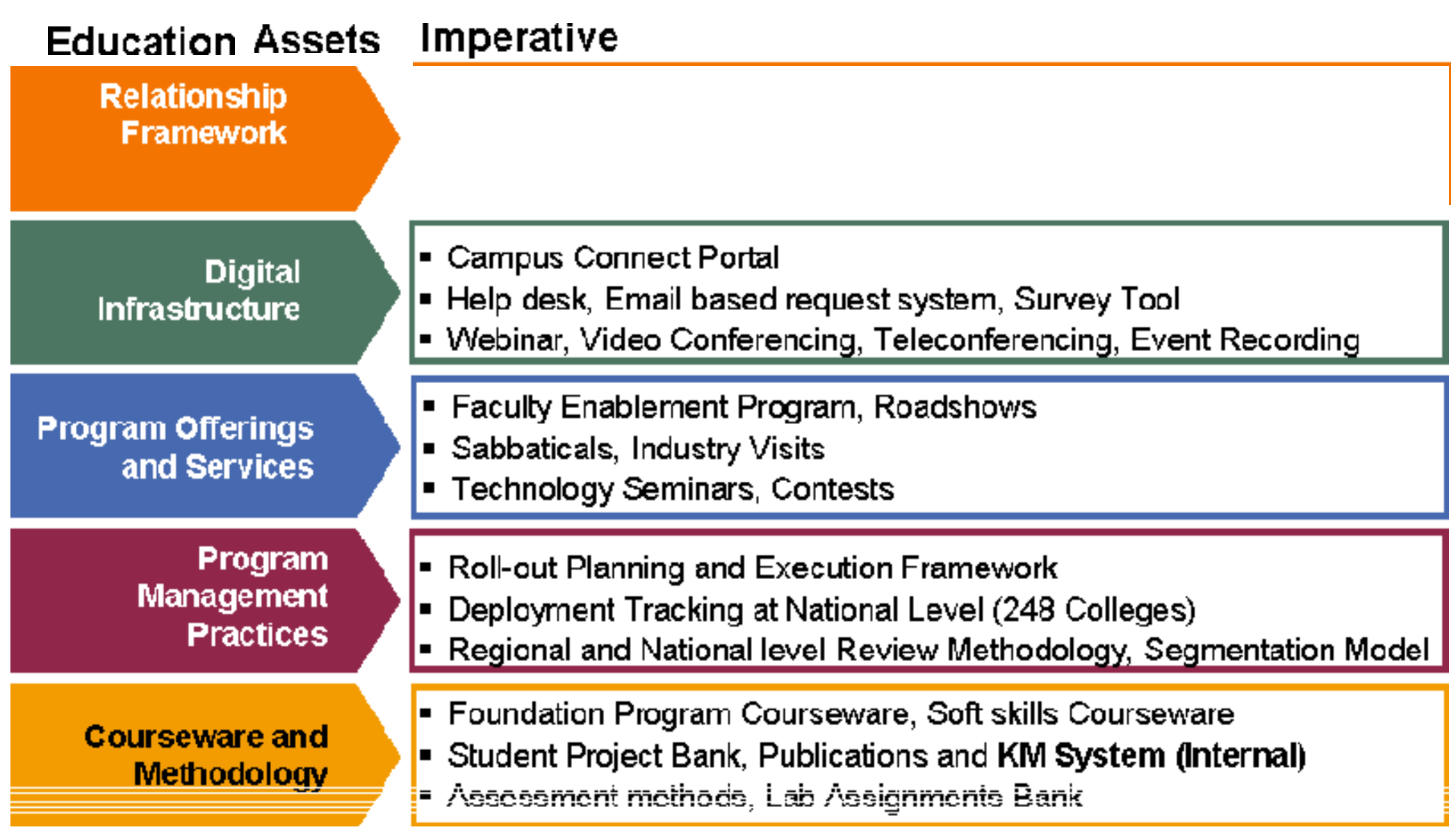

**Fig 2 – Educationware [2]**

Other Corporates in the Industry should be encouraged to adopt similar models. The corporate-academia partnership has four advantages: it ensures that the content for the programmes is relevant and well-timed, is delivered at world-class standards, is global in its perspectives and finally it helps in brand building of the company. Industry gains from getting better equipped software engineers who have been trained by faculty and are imbibed with skills from the Industry.

## 5.2 Industry internships and live projects to students in the final year of Engineering

Software Companies and Corporates must provide internships and projects to final year engineering students. This will help in giving a rounding off to the education the students have received till then.

Exposure to real-life projects will help prepare software engineers who can better face the challenges of the industry. This would stand in good stead in comparison to the dummy projects that are presently done in various institutions. Also it would be a good idea to involve students in industry projects for a period longer than one semester [9]. Most companies offer 3 month and 6 month projects for internship for students of bachelor's and master's degrees.

### 5.3 Training of newly recruited software engineers to include more relevant subjects/topics

The training deployed for newly recruited employees should include more relevant topics such as Project Management Fundamentals, Requirements engineering and Management, Data Modeling, Process Modeling, System Development Life Cycle concepts, Maintenance and Off-shoring concepts, etc. It would be advantageous also to include project work or time in a project as part of the training. If this is done employees can apply their newly acquired skills and determine their validity [3].

## 6 CONCLUSION

Industry has a very important role to play in the shaping of the future of their employees and consequently the nation. Software companies and corporates have to take a lead in helping academia to shape the curriculum so that software engineers who can tackle the industry challenges are prepared. Academia on the other hand should analyse and incorporate changes constantly so that technology changes can be reflected as required. Both universities and institutions on one hand and companies on the other hand have to work in tandem to ensure the grooming of students who will be software engineers of tomorrow. Industry would be the first to gain from helping shape future employees as they need to endure less training costs and would get well equipped software engineers. Academia would learn and be able to better teach their pupils thus discharging their duties in an improved manner.